\documentclass[proof]{WileyASNA-v1}
\UseRawInputEncoding

\articletype{Proceedings}%

\received{}
\revised{}
\accepted{}
\newcommand{\Msun}{M_\odot}
\begin{document}

\title{Counting States: A Combinatorial Analysis of SQM Fragmentation}

\author[1]{A. Bernardo*}

\author[2]{L. Paulucci}

\author[1]{L. M. de S\'{a}}

\author[1]{J.E. Horvath}
\authormark{A. Bernardo \textsc{et al}}

\address[1]{\orgdiv{Universidade de S\~{a}o Paulo}, \orgname{Instituto de Astronomia, Geof\'{i}sica e Ci\^{e}ncias Atmosf\'{e}ricas}, \orgaddress{\city{S\~{a}o Paulo}, \state{SP}, \country{Brazil}}}

\address[2]{\orgname{UFABC}, \orgaddress{\city{Santo Andr\'{e}} \country{Brazil}}}

\corres{*Antonio Bernardo, \email{Alc.bernardo@gmail.com}}

\presentaddress{R. do Mat\~{a}o, 1226 - Cidade Universit\'{a}ria, 05508-090, S\~{a}o Paulo-SP, Brazil}

\abstract{The Strange Quark matter (SQM) hypothesis states that at extreme pressure and density conditions a new ground state of matter would arise, in which half of the \textit{down} quarks become strange quarks. If true, it would mean that at least the core of neutron stars is made of SQM. In this hypothesis, SQM would be released in the inter-stellar medium when two of these objects merge. It is estimated that $10^{-2} \Msun$ of SQM would be released this way. This matter will undergo a sequence of processes that should result in a fraction of the released SQM becoming heavy nuclei through \textit{r-process}. In this work we are interested in characterizing the fragmentation of SQM, with the novelty of keeping track of the \textit{quark configuration} of the fragmented matter. This is accomplished by developing a methodology to estimate the energy of each fragment as the sum of its \textit{constituent quarks}, the Coulomb interaction among the quarks and fragments' momenta.}


\jnlcitation{\cname{%
\author{A. Bernardo }, 
\author{L. Paulucci}, 
\author{L. M. de S\'{a}}, and 
\author{J. E. Horvath }} (\cyear{2016}), 
\ctitle{Counting States: A Combinatorial Analysis of SQM Fragmentation}, \cjournal{Astron. Nachrichten}, \cvol{Unknown}.}

\maketitle

\section{Introduction}\label{sec1}
Understanding the origin and abundances of heavy elements in the Universe is a great challenge. Recent observations of NS-NS mergers suggests that the second and third peak of r-process elements are mainly synthesized in these events \citep{Cote2018,Chornock2017}.

In this work we work with the strange hypothesis from \cite{Witten1984}, which states that under the conditions in the interior of NS the preferred state of matter is \textit{strange quark matter} (SQM) \citep{Benvenuto1989, Rocha2020}. The merger of strange stars has been explored by \cite{Rosinka2008,Pietri2019}, and the process which the ejecta released at the merger undergoes can be broken into stages according to their time scales.
\begin{table*}[hb!]%
\caption{This chart shows the three  partitions of A=3\label{tab1}}
\centering
\begin{tabular}{|lllllllll|llllll|lll|}
\hline
\multicolumn{9}{|c|}{$\lambda_1= [3,0,0]$}                                                              & \multicolumn{6}{c|}{$\lambda_2=[1,1,0]$}      & \multicolumn{3}{c|}{$\lambda_3=[0,0,1]$}                       \\ \hline
\multicolumn{3}{|c|}{$i=1$}               & \multicolumn{3}{c|}{$i=1$}                & \multicolumn{3}{c|}{$i=1$} & \multicolumn{3}{c|}{$i=1$}        & \multicolumn{3}{c|}{$i=2$ }        & \multicolumn{3}{c|}{$i=3$}\\ \hline
$N_u$ & $N_d$ & \multicolumn{1}{l|}{$N_s$} & $N_u$ & $N_d$ & \multicolumn{1}{l|}{$N_s$} & $N_u$   & $N_d$   & $N_s$   & $N_u$ & $N_d$ & \multicolumn{1}{l|}{$N_s$} & $N_u$   & $N_d$   & \multicolumn{1}{l|}{$N_s$ }  & $N_u$   & $N_d$   & \multicolumn{1}{l|}{$N_s$ }  \\ \hline
3    & 0    & \multicolumn{1}{l|}{0}    & 0    & 3    & \multicolumn{1}{l|}{0}    & 0      & 0      & 3      & 3    & 0    & \multicolumn{1}{l|}{0}    & 0      & 3      & 3      & 3     & 3      & 3    \\
3    & 0    & \multicolumn{1}{l|}{0}    & 0    & 2    & \multicolumn{1}{l|}{1}    & 0      & 1      & 2      & 2    & 1    & \multicolumn{1}{l|}{0}    & 1      & 2      & 3  & - &- &\multicolumn{1}{l|}{-} \\
2    & 1    & \multicolumn{1}{l|}{0}    & 1    & 2    & \multicolumn{1}{l|}{0}    & 0      & 0      & 3      & 1    & 2    & \multicolumn{1}{l|}{0}    & 2      & 1      & 3  &- &- &\multicolumn{1}{l|}{-}  \\
2    & 1    & \multicolumn{1}{l|}{0}    & 1    & 1    & \multicolumn{1}{l|}{1}    & 0      & 1      & 2      & 0    & 3    & \multicolumn{1}{l|}{0}    & 3      & 0      & 3   & -& -&\multicolumn{1}{l|}{-} \\
2    & 1    & \multicolumn{1}{l|}{0}    & 0    & 2    & \multicolumn{1}{l|}{1}    & 1      & 0      & 2      & 2    & 0    & \multicolumn{1}{l|}{1}    & 1      & 3      & 2   & -&-&\multicolumn{1}{l|}{-} \\
1    & 2    & \multicolumn{1}{l|}{0}    & 2    & 0    & \multicolumn{1}{l|}{1}    & 0      & 1      & 2      & 1    & 1    & \multicolumn{1}{l|}{1}    & 2      & 2      & 2  & -& -&\multicolumn{1}{l|}{-} \\
1    & 2    & \multicolumn{1}{l|}{0}    & 1    & 1    & \multicolumn{1}{l|}{1}    & 1      & 0      & 2      & 0    & 2    & \multicolumn{1}{l|}{1}    & 3      & 1      & 2   & -& -&\multicolumn{1}{l|}{-} \\
0    & 3    & \multicolumn{1}{l|}{0}    & 2    & 0    & \multicolumn{1}{l|}{1}    & 1      & 0      & 2      & 1    & 0    & \multicolumn{1}{l|}{2}    & 2      & 3      & 1   & -& -&\multicolumn{1}{l|}{-} \\
2    & 0    & \multicolumn{1}{l|}{1}    & 1    & 1    & \multicolumn{1}{l|}{1}    & 0      & 2      & 1      & 0    & 1    & \multicolumn{1}{l|}{2}    & 3      & 2      & 1  &- &- &\multicolumn{1}{l|}{-} \\
1    & 1    & \multicolumn{1}{l|}{1}    & 1    & 1    & \multicolumn{1}{l|}{1}    & 1      & 1      & 1      & 0    & 0    & \multicolumn{1}{l|}{3}    & 3      & 3      & 0  &- & -&\multicolumn{1}{l|}{-} \\ \hline
\end{tabular}
\end{table*}

The first stage (fragmentation/hadronization) happens on a strong interaction time scale $\tau_s=10^{-24}s$. The compressed quark matter is released from the stars into the adjacent interstellar medium where it cools and expands. The change of the thermodynamic quantities forces the system into a new state of maximum entropy in which a portion of the released quark matter will become hadrons of various baryonic numbers \citep{Bugaev2000,Paulucci2014,Bucciantini2019}; while the remaining matter will remain as quark matter and roam the interstellar medium \citep{Horvath2008,Lugones2004}, possibly serving as seeds for \textit{strangefication} of other neutron stars \citep{Benvenuto1989B}.

The following stages are the decay and recombination, they happen concomitantly at a weak interaction time scale $\tau_w=10^{-14}s$. The metastable baryons that arise from the former stage break into \textit{neutrons} and \textit{protons} releasing mesons in the process. The mesons will majorly decay into \textit{muons} that can be captured by the protons, with a cross sections larger than the usual beta-inverse process due to the relatively large $\mu^-$ mass. As the ejecta cools and expands these reactions freeze, setting a $Y_e$ - that represents the number of protons by the number of nucleons - for the next stage, the nucleosynthesis.

The nucleosynthesis will take place by means of \textit{r-process} \citep{Freiburghaus1999, Paulucci2017, Horvath2019}. The abundance of each element will depend primarily on the $Y_e$ that is defined by all of the previous stages. In this work we further investigate the fragmentation process.

The formalism to be introduced in this work is reminiscent of the old \textit{Quark Combinatorics} \citep{Valoshin1982,Guman1975,Anisovich1978}. We assume that the SQM is constituted by equal amounts of \textit{u, d} \& \textit{s} quarks, that remain unchanged during the fragmentation stage due to $\tau_s << \tau_w$. The probability of a given fragmentation spectrum ($\lambda$) to be chosen by the system is proportional to the number of ways in which this spectrum can be attained while independently conserving the number of each quark flavor and maximizing the phase space volume reachable by the system. The remainder of this paper tries to elucidate how to calculate those quantities.
\section{Methodology}\label{sec2}
\subsection{Partitions}
The first step to characterize the fragmentation of a chunk of SQM of baryonic number $A$ is to find which of the $\pi(A)$ partitions maximizes the system's entropy. For large A we have:
\begin{equation*}
\pi(A) \sim \frac{1}{4A\sqrt{3}}exp\left(\pi\sqrt{\frac{2A}{3}}\right)
\end{equation*}  

We use the partition function $Z(\lambda)$ where $\lambda$ is one partition (or baryonic spectrum) of A, expressed as $\lambda_j=[n_1,n_2,...,n_A]$. The $i^{th}$ entry counts the number of fragments of size \textit{i} respecting:
\begin{equation*}
    A=\sum_{i=1}^A i n_i
\end{equation*}
The Partition function is given by the expression:
\begin{equation*}
Z(\lambda)=\Omega(\lambda)\sum_{i=1}^{M(\lambda)} \left( \frac{\pi m_i}{2\beta}\right)^{3/2} e^{-\beta m_i} ,
\end{equation*}
where $M(\lambda)$ is the number of fragments, $m_i$ the mass/energy of each fragment and $\Omega(\lambda)$ is the number of configurations of $\lambda$. 

We are only considering the configurations $\nu$ of $\lambda$ that observe the condition \textbf{$N_u=N_d=N_s=A$}. Where $N_u$, $N_d$, $N_s$ are the number of quarks \textit{up, down} and \textit{strange} respectively. Such constrain on the configurations makes it very difficult to calculate.
Without the constrain $\Omega(\lambda)$ would simply be:

\begin{equation*}
\Bar{\Omega}(\lambda)=\prod_i^{M(\lambda)}\binom{\frac{9}{2}(i^2 +i) + n_i}{\frac{9}{2}(i^2 +i)}
\end{equation*}

With the constrains we have to apply Polya's enumeration theorem to arrive at the following structure:
\begin{equation*} 
\Omega(\lambda)=\left[(uds)^A\right]
\prod_{i=1}^{A}
\mathcal{Z}\left(S_{n_i}; 
\mathcal{Z}\left(S_i; u + d + s\right)\right)
\end{equation*}
$S_i$ is the symetric group of order $i$, and $\mathcal{Z}$ is the cycle index polynomial. 
It is very computationally expensive to calculate $\Omega(\lambda)$ using this expression. We are working on getting an analytical function to represent $\Omega(\lambda)$ to enable calculations of order $A\sim 10^{40}$.
\subsection{Configurations}

Once the partition is determined it is possible to choose the configuration $\nu$ of maximum entropy.

A configuration is a way to distribute A quarks of each type into a partition $\lambda$.
For instance, $A=3$ has 3 partitions: $\lambda_1= [3,0,0]$, $\lambda_2 = [1,1,0]$ and $\lambda_3=[0,0,1]$, we have that $\Omega(\lambda_1)=\Omega(\lambda_2)=10$ and $\Omega(\lambda_3)=1$. The configurations of each partition of 3 can be found on Table \ref{tab1}

The partition function of each \textit{configuration} will be given by:
\begin{equation*}
Z(\nu)=\sum_i^{M(\lambda)} \left( \frac{\pi m_i(\nu)}{2\beta}\right)^{3/2} e^{-\beta m_i(\nu)} 
\end{equation*}

Once the configuration of maximal entropy is found we can consider the fragmentation output fully determined. $\lambda$ shows the baryonic spectrum and $\nu$ shows the abundance of each baryon and how much of the SQM will remain as strangelets.
\subsection{Fragments Energy}
To apply the formalism above we have to estimate the mass of each fragment, depending both of the partition and the configuration.
To do this, we sum the constituent masses of each quark \citep{Yang2020} and Coulombian contributions.   
\begin{equation*}
m_{i}=N_{u}m_{u}+N_{d}m_{d}+N_{s}m_{s} +k_e \frac{(q^2/2-q/6-A/9)}{r_o A^{1/3}}
\end{equation*}
$q=(\frac{2N_{u}}{3}-\frac{N_{d}}{2}-\frac{N_{s}}{2}) e^-$ is the strangelet electric charge, $k_e$ the coulomb constant, $r_o$ the radius of a strangelet of $A=1$ and $m_u=313$ MeV, $m_d=313$  MeV, and $m_s=555$ MeV are the constituent masses of each quark. We intend to use more intricate masses estimations in the future, taking quark momenta and degeneracy into account, as well as CFLs variations \citep{Horvath2003}.
\section{Results}
As an example and proof of concept, we will carry the calculations for the simple case of $A=6$.
\begin{center}   
\begin{table}[h!]
\centering
\caption{Partitions of A=6, number of configuration and fragments\label{tab2}}
\begin{tabular}{llllllcc}
\multicolumn{6}{c}{$\lambda$} & $\Omega(\lambda)$ & $\textit{M}(\lambda)$ \\
{[}6,  & 0,  & 0,  & 0, & 0, & 0{]} & 103 & 6                                  \\
{[}4,  & 1,  & 0,  & 0, & 0, & 0{]} & 418 & 5                                  \\
{[}3,  & 0,  & 1,  & 0, & 0, & 0{]} & 193 & 4                                  \\
{[}2,  & 2,  & 0,  & 0, & 0, & 0{]} & 442 & 4                                  \\
{[}2,  & 0,  & 0,  & 1, & 0, & 0{]} & 55  & 3                                  \\
{[}1,  & 1,  & 1,  & 0, & 0, & 0{]} & 235 & 3                                  \\
{[}1,  & 0,  & 0,  & 0, & 1, & 0{]} & 10  & 2                                  \\
{[}0,  & 3,  & 0,  & 0, & 0, & 0{]} & 73  & 3                                  \\
{[}0,  & 1,  & 0,  & 1, & 0, & 0{]} & 28  & 2                                  \\
{[}0,  & 0,  & 2,  & 0, & 0, & 0{]} & 19  & 2                                  \\
{[}0,  & 0,  & 0,  & 0, & 0, & 1{]} & 1   & 1                              
\end{tabular}
\end{table}
\end{center}
  Figure \ref{fig1} shows the probability of each partition being the result of the fragmentation, with temperatures of 2 MeV, 5 MeV and 50 MeV, represented by the colors red, blue and black respectively. The preferred partition $\lambda=[4,1,0,0,0,0]$ is the same for any temperature of interest, it is noticeable that the probabilities barely change with the temperature. The choice for this particular $\lambda$ results from a big $\Omega(\lambda)$ and lots of small fragments.
  
  Figure \ref{fig2} shows the probability of each configuration (x-axes, in a defined ordering), changing the temperature. The most probable configuration for $A=6$ is 2 protons, 2 neutrons and all the 6 s-quarks in a single hadron.

\begin{center}
\includegraphics[width=0.9\linewidth]{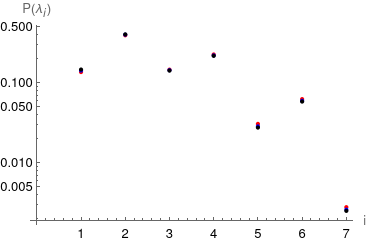}
\captionof{figure}{Probability of each partition with A=6, changing the temperature. The x-axes represents the index of each partition, in the same ordering of Table \ref{tab2} \label{fig1}}
\end{center}

This kind of \textit{distillation} is caused by the system maximizing the phase space volume (and therefore, the entropy) by creating the maximum number of light fragments possible, while maintaining all the \textit{s-quarks} on the minimum number o fragments.

\begin{center}
\includegraphics[width=0.9\linewidth]{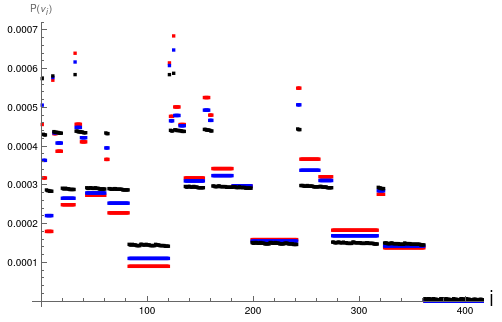}
\captionof{figure}{Probability of each configuration for the second partition of A=6, with temperatures of 2 MeV, 5 MeV and 50 MeV, represented by the colors red, blue and black respectively\label{fig2}}
\end{center}
One could expect \textit{all-$\Lambda$s} to be the product of fragmentation \citep{Bethe1987}. However, if the results we obtained for small \textit{A} hold true for realistic quantities of matter, the \textit{all-$\Lambda$s} output will be severely suppressed in favor of the most distilled configurations.
\section{Conclusions}\label{sec5}

We showed how the output of the fragmentation process can change by enumerating all the possible outcomes that observe the conservation rules expected for the system. This formalism has difficulties due to its complex combinatorial nature, but asymptotic relations are being developed.

As a rule, we can expect distilled configurations to be the most probable, with a negligible dependence on the temperature. The evolution of hadrons output with $A>1$ is yet to be studied, some of them will stay as strange nuggets and roam the interstellar medium, while the less massive ones should produce ordinary hadrons and participate on the nucleosynthesis. 

Depending on the resulting configuration, $Y_e$ can vary drastically.  The determination of the fragmentation output is crucial to fully characterize the subsequent nucleosynthesis. For example, a canonical all $\Lambda$s fragmentation will yield an electron fraction of $Y_e= 2/3$ that results in a weak \textit{r-process}. Half $\Sigma^-$(dds) and half $\Sigma^+$(uus), a configuration that also has equal abundance of the three quarks, would yield a $Y_e=1/4$ that produces heavier elements.


\section*{Acknowledgments}
Financial support was provided by the Fapesp Agency, Capes and CNPq
\bibliography{Wiley-ASNA}%

\begin{thebibliography}{}

\bibitem [\protect \citeauthoryear {%
Anisovich%
}{%
Anisovich%
}{%
{\protect \APACyear {1978}}%
}]{%
Anisovich1978}
\APACinsertmetastar {%
Anisovich1978}%
\begin{APACrefauthors}%
Anisovich, V\BPBI V.%
\end{APACrefauthors}%
\unskip\
\newblock
\APACrefYearMonthDay{1978}{9}{},
\newblock
\unskip
\newblock
\APACjournalVolNumPages{Sov. J. Nucl. Phys. (Engl. Transl.); (United
  States)}{28:3}{}{}.
\newblock
\begin{APACrefURL} \url{https://www.osti.gov/biblio/6249029} \end{APACrefURL}
\PrintBackRefs{\CurrentBib}

\bibitem [\protect \citeauthoryear {%
Benvenuto%
\ \BBA {} Horvath%
}{%
Benvenuto%
\ \BBA {} Horvath%
}{%
{\protect \APACyear {1989}}%
{\protect \APACexlab {{\protect \BCnt {1}}}}}]{%
Benvenuto1989}
\APACinsertmetastar {%
Benvenuto1989}%
\begin{APACrefauthors}%
Benvenuto, O\BPBI G.%
\BCBT {}\ \BBA {} Horvath, J\BPBI E.%
\end{APACrefauthors}%
\unskip\
\newblock
\APACrefYearMonthDay{1989{\protect \BCnt {1}}}{Aug}{},
\newblock
\unskip
\newblock
\APACjournalVolNumPages{Phys. Rev. Lett.}{63}{}{716--719}.
\newblock
\begin{APACrefURL} \url{https://link.aps.org/doi/10.1103/PhysRevLett.63.716}
  \end{APACrefURL}
\PrintBackRefs{\CurrentBib}

\bibitem [\protect \citeauthoryear {%
Benvenuto%
\ \BBA {} Horvath%
}{%
Benvenuto%
\ \BBA {} Horvath%
}{%
{\protect \APACyear {1989}}%
{\protect \APACexlab {{\protect \BCnt {2}}}}}]{%
Benvenuto1989B}
\APACinsertmetastar {%
Benvenuto1989B}%
\begin{APACrefauthors}%
Benvenuto, O\BPBI G.%
\BCBT {}\ \BBA {} Horvath, J\BPBI E.%
\end{APACrefauthors}%
\unskip\
\newblock
\APACrefYearMonthDay{1989{\protect \BCnt {2}}}{{\APACmonth{06}}}{},
\newblock
\unskip
\newblock
\APACjournalVolNumPages{Mod. Phys. Lett. A}{04}{12}{1085--1089}.
\newblock
\begin{APACrefURL} \url{https://doi.org/10.1142/s021773238900126x}
  \end{APACrefURL}
\PrintBackRefs{\CurrentBib}

\bibitem [\protect \citeauthoryear {%
Bethe%
, Brown%
\BCBL {}\ \BBA {} Cooperstein%
}{%
Bethe%
\ \protect \BOthers {.}}{%
{\protect \APACyear {1987}}%
}]{%
Bethe1987}
\APACinsertmetastar {%
Bethe1987}%
\begin{APACrefauthors}%
Bethe, H.%
, Brown, G.%
\BCBL {}\ \BBA {} Cooperstein, J.%
\end{APACrefauthors}%
\unskip\
\newblock
\APACrefYearMonthDay{1987}{}{},
\newblock
\unskip
\newblock
\APACjournalVolNumPages{Nucl. Phys. A}{462}{4}{791-802}.
\newblock
\begin{APACrefURL}
  \url{https://www.sciencedirect.com/science/article/pii/037594748790577X}
  \end{APACrefURL}
\PrintBackRefs{\CurrentBib}

\bibitem [\protect \citeauthoryear {%
Bucciantini%
, Drago%
, Pagliara%
\BCBL {}\ \BBA {} Traversi%
}{%
Bucciantini%
\ \protect \BOthers {.}}{%
{\protect \APACyear {2019}}%
}]{%
Bucciantini2019}
\APACinsertmetastar {%
Bucciantini2019}%
\begin{APACrefauthors}%
Bucciantini, N.%
, Drago, A.%
, Pagliara, G.%
\BCBL {}\ \BBA {} Traversi, S.%
\end{APACrefauthors}%
\unskip\
\newblock
\APACrefYearMonthDay{2019}{}{},
\newblock
\APACrefbtitle {Formation and evaporation of strangelets during the merger of
  two compact stars.} {Formation and evaporation of strangelets during the
  merger of two compact stars.}
\newblock
\begin{APACrefURL} \url{https://arxiv.org/abs/1908.02501} \end{APACrefURL}
\newblock
\begin{APACrefDOI} \doi{10.48550/ARXIV.1908.02501} \end{APACrefDOI}
\PrintBackRefs{\CurrentBib}

\bibitem [\protect \citeauthoryear {%
Bugaev%
, Gorenstein%
, Mishustin%
\BCBL {}\ \BBA {} Greiner%
}{%
Bugaev%
\ \protect \BOthers {.}}{%
{\protect \APACyear {2000}}%
}]{%
Bugaev2000}
\APACinsertmetastar {%
Bugaev2000}%
\begin{APACrefauthors}%
Bugaev, K\BPBI A.%
, Gorenstein, M\BPBI I.%
, Mishustin, I\BPBI N.%
\BCBL {}\ \BBA {} Greiner, W.%
\end{APACrefauthors}%
\unskip\
\newblock
\APACrefYearMonthDay{2000}{Sep}{},
\newblock
\unskip
\newblock
\APACjournalVolNumPages{Phys. Rev. C}{62}{}{044320}.
\newblock
\begin{APACrefURL} \url{https://link.aps.org/doi/10.1103/PhysRevC.62.044320}
  \end{APACrefURL}
\PrintBackRefs{\CurrentBib}

\bibitem [\protect \citeauthoryear {%
Chornock%
\ \protect \BOthers {.}}{%
Chornock%
\ \protect \BOthers {.}}{%
{\protect \APACyear {2017}}%
}]{%
Chornock2017}
\APACinsertmetastar {%
Chornock2017}%
\begin{APACrefauthors}%
Chornock, R.%
, Berger, E.%
, Kasen, D.%
\ et al.\end{APACrefauthors}%
\unskip\
\newblock
\APACrefYearMonthDay{2017}{oct}{},
\newblock
\unskip
\newblock
\APACjournalVolNumPages{The Astrophysical Journal Letters}{848}{2}{L19}.
\newblock
\begin{APACrefURL} \url{https://dx.doi.org/10.3847/2041-8213/aa905c}
  \end{APACrefURL}
\PrintBackRefs{\CurrentBib}

\bibitem [\protect \citeauthoryear {%
C\^ot\'e%
\ \protect \BOthers {.}}{%
C\^ot\'e%
\ \protect \BOthers {.}}{%
{\protect \APACyear {2018}}%
}]{%
Cote2018}
\APACinsertmetastar {%
Cote2018}%
\begin{APACrefauthors}%
C\^ot\'e, B.%
, Fryer, C\BPBI L.%
, Belczynski, K.%
\ et al.\end{APACrefauthors}%
\unskip\
\newblock
\APACrefYearMonthDay{2018}{mar}{},
\newblock
\unskip
\newblock
\APACjournalVolNumPages{The Astrophysical Journal}{855}{2}{99}.
\newblock
\begin{APACrefURL} \url{https://dx.doi.org/10.3847/1538-4357/aaad67}
  \end{APACrefURL}
\PrintBackRefs{\CurrentBib}

\bibitem [\protect \citeauthoryear {%
Freiburghaus%
, Rosswog%
\BCBL {}\ \BBA {} Thielemann%
}{%
Freiburghaus%
\ \protect \BOthers {.}}{%
{\protect \APACyear {1999}}%
}]{%
Freiburghaus1999}
\APACinsertmetastar {%
Freiburghaus1999}%
\begin{APACrefauthors}%
Freiburghaus, C.%
, Rosswog, S.%
\BCBL {}\ \BBA {} Thielemann, F\BHBI K.%
\end{APACrefauthors}%
\unskip\
\newblock
\APACrefYearMonthDay{1999}{oct}{},
\newblock
\unskip
\newblock
\APACjournalVolNumPages{The Astrophysical Journal}{525}{2}{L121}.
\newblock
\begin{APACrefURL} \url{https://dx.doi.org/10.1086/312343} \end{APACrefURL}
\PrintBackRefs{\CurrentBib}

\bibitem [\protect \citeauthoryear {%
Gondek-Rosinska%
\ \BBA {} Limousin%
}{%
Gondek-Rosinska%
\ \BBA {} Limousin%
}{%
{\protect \APACyear {2008}}%
}]{%
Rosinka2008}
\APACinsertmetastar {%
Rosinka2008}%
\begin{APACrefauthors}%
Gondek-Rosinska, D.%
\BCBT {}\ \BBA {} Limousin, F.%
\end{APACrefauthors}%
\unskip\
\newblock
\APACrefYearMonthDay{2008}{}{},
\newblock
\APACrefbtitle {The final phase of inspiral of strange quark star binaries.}
  {The final phase of inspiral of strange quark star binaries.}
\newblock
\begin{APACrefURL} \url{https://arxiv.org/abs/0801.4829} \end{APACrefURL}
\PrintBackRefs{\CurrentBib}

\bibitem [\protect \citeauthoryear {%
Guman%
\ \BBA {} Shekhter%
}{%
Guman%
\ \BBA {} Shekhter%
}{%
{\protect \APACyear {1975}}%
}]{%
Guman1975}
\APACinsertmetastar {%
Guman1975}%
\begin{APACrefauthors}%
Guman, V.%
\BCBT {}\ \BBA {} Shekhter, V.%
\end{APACrefauthors}%
\unskip\
\newblock
\APACrefYearMonthDay{1975}{}{},
\newblock
\unskip
\newblock
\APACjournalVolNumPages{Nucl. Phys. B}{99}{3}{523-540}.
\newblock
\begin{APACrefURL}
  \url{https://www.sciencedirect.com/science/article/pii/S0550321375800265}
  \end{APACrefURL}
\PrintBackRefs{\CurrentBib}

\bibitem [\protect \citeauthoryear {%
Horvath%
}{%
Horvath%
}{%
{\protect \APACyear {2008}}%
}]{%
Horvath2008}
\APACinsertmetastar {%
Horvath2008}%
\begin{APACrefauthors}%
Horvath, J\BPBI E.%
\end{APACrefauthors}%
\unskip\
\newblock
\APACrefYearMonthDay{2008}{{\APACmonth{04}}}{},
\newblock
\unskip
\newblock
\APACjournalVolNumPages{Astrophysics and Space Science}{315}{1-4}{361--364}.
\newblock
\begin{APACrefURL} \url{https://doi.org/10.1007/s10509-008-9783-x}
  \end{APACrefURL}
\PrintBackRefs{\CurrentBib}

\bibitem [\protect \citeauthoryear {%
Horvath%
\ \protect \BOthers {.}}{%
Horvath%
\ \protect \BOthers {.}}{%
{\protect \APACyear {2019}}%
}]{%
Horvath2019}
\APACinsertmetastar {%
Horvath2019}%
\begin{APACrefauthors}%
Horvath, J\BPBI E.%
, Benvenuto, O\BPBI G.%
, Bauer, E.%
, Paulucci, L.%
, Bernardo, A.%
\BCBL {}\ \BBA {} Viturro, H\BPBI R.%
\end{APACrefauthors}%
\unskip\
\newblock
\APACrefYearMonthDay{2019}{{\APACmonth{06}}}{},
\newblock
\unskip
\newblock
\APACjournalVolNumPages{Universe}{5}{6}{144}.
\newblock
\begin{APACrefURL} \url{https://doi.org/10.3390/universe5060144}
  \end{APACrefURL}
\PrintBackRefs{\CurrentBib}

\bibitem [\protect \citeauthoryear {%
Horvath%
, Lugones%
\BCBL {}\ \BBA {} De~Freitas~Pacheco%
}{%
Horvath%
\ \protect \BOthers {.}}{%
{\protect \APACyear {2003}}%
}]{%
Horvath2003}
\APACinsertmetastar {%
Horvath2003}%
\begin{APACrefauthors}%
Horvath, J\BPBI E.%
, Lugones, G.%
\BCBL {}\ \BBA {} De~Freitas~Pacheco, J\BPBI A.%
\end{APACrefauthors}%
\unskip\
\newblock
\APACrefYearMonthDay{2003}{}{},
\newblock
\unskip
\newblock
\APACjournalVolNumPages{IJMPD}{12}{03}{519-526}.
\newblock
\begin{APACrefURL} \url{https://doi.org/10.1142/S0218271803002743}
  \end{APACrefURL}
\PrintBackRefs{\CurrentBib}

\bibitem [\protect \citeauthoryear {%
Lugones%
\ \BBA {} Horvath%
}{%
Lugones%
\ \BBA {} Horvath%
}{%
{\protect \APACyear {2004}}%
}]{%
Lugones2004}
\APACinsertmetastar {%
Lugones2004}%
\begin{APACrefauthors}%
Lugones, G.%
\BCBT {}\ \BBA {} Horvath, J\BPBI E.%
\end{APACrefauthors}%
\unskip\
\newblock
\APACrefYearMonthDay{2004}{{\APACmonth{03}}}{},
\newblock
\unskip
\newblock
\APACjournalVolNumPages{PRD}{69}{6}{}.
\newblock
\begin{APACrefURL} \url{https://doi.org/10.1103/physrevd.69.063509}
  \end{APACrefURL}
\PrintBackRefs{\CurrentBib}

\bibitem [\protect \citeauthoryear {%
Paulucci%
\ \BBA {} Horvath%
}{%
Paulucci%
\ \BBA {} Horvath%
}{%
{\protect \APACyear {2014}}%
}]{%
Paulucci2014}
\APACinsertmetastar {%
Paulucci2014}%
\begin{APACrefauthors}%
Paulucci, L.%
\BCBT {}\ \BBA {} Horvath, J.%
\end{APACrefauthors}%
\unskip\
\newblock
\APACrefYearMonthDay{2014}{}{},
\newblock
\unskip
\newblock
\APACjournalVolNumPages{Physics Letters B}{733}{}{164-168}.
\newblock
\begin{APACrefURL}
  \url{https://www.sciencedirect.com/science/article/pii/S0370269314002767}
  \end{APACrefURL}
\PrintBackRefs{\CurrentBib}

\bibitem [\protect \citeauthoryear {%
Paulucci%
, Horvath%
\BCBL {}\ \BBA {} Benvenuto%
}{%
Paulucci%
\ \protect \BOthers {.}}{%
{\protect \APACyear {2017}}%
}]{%
Paulucci2017}
\APACinsertmetastar {%
Paulucci2017}%
\begin{APACrefauthors}%
Paulucci, L.%
, Horvath, J\BPBI E.%
\BCBL {}\ \BBA {} Benvenuto, O.%
\end{APACrefauthors}%
\unskip\
\newblock
\APACrefYearMonthDay{2017}{}{},
\newblock
\unskip
\newblock
\APACjournalVolNumPages{International Journal of Modern Physics: Conference
  Series}{45}{}{}.
\newblock
\begin{APACrefURL} \url{https://doi.org/10.1142/S2010194517600424}
  \end{APACrefURL}
\PrintBackRefs{\CurrentBib}

\bibitem [\protect \citeauthoryear {%
Pietri%
\ \protect \BOthers {.}}{%
Pietri%
\ \protect \BOthers {.}}{%
{\protect \APACyear {2019}}%
}]{%
Pietri2019}
\APACinsertmetastar {%
Pietri2019}%
\begin{APACrefauthors}%
Pietri, R\BPBI D.%
, Drago, A.%
, Feo, A.%
, Pagliara, G.%
, Pasquali, M.%
, Traversi, S.%
\BCBL {}\ \BBA {} Wiktorowicz, G.%
\end{APACrefauthors}%
\unskip\
\newblock
\APACrefYearMonthDay{2019}{aug}{},
\newblock
\unskip
\newblock
\APACjournalVolNumPages{The Astrophysical Journal}{881}{2}{122}.
\newblock
\begin{APACrefURL} \url{https://dx.doi.org/10.3847/1538-4357/ab2fd0}
  \end{APACrefURL}
\PrintBackRefs{\CurrentBib}

\bibitem [\protect \citeauthoryear {%
Rocha%
, Bernardo%
, Avellar%
\BCBL {}\ \BBA {} Horvath%
}{%
Rocha%
\ \protect \BOthers {.}}{%
{\protect \APACyear {2020}}%
}]{%
Rocha2020}
\APACinsertmetastar {%
Rocha2020}%
\begin{APACrefauthors}%
Rocha, L\BPBI S.%
, Bernardo, A.%
, Avellar, M\BPBI G\BPBI B\BPBI D.%
\BCBL {}\ \BBA {} Horvath, J\BPBI E.%
\end{APACrefauthors}%
\unskip\
\newblock
\APACrefYearMonthDay{2020}{{\APACmonth{05}}}{},
\newblock
\unskip
\newblock
\APACjournalVolNumPages{IJMPD}{29}{07}{2050044}.
\newblock
\begin{APACrefURL} \url{https://doi.org/10.1142/s0218271820500443}
  \end{APACrefURL}
\PrintBackRefs{\CurrentBib}

\bibitem [\protect \citeauthoryear {%
Voloshin%
, Nikitin%
\BCBL {}\ \BBA {} Porfirov%
}{%
Voloshin%
\ \protect \BOthers {.}}{%
{\protect \APACyear {1982}}%
}]{%
Valoshin1982}
\APACinsertmetastar {%
Valoshin1982}%
\begin{APACrefauthors}%
Voloshin, S\BPBI A.%
, Nikitin, Y\BPBI P.%
\BCBL {}\ \BBA {} Porfirov, P\BPBI I.%
\end{APACrefauthors}%
\unskip\
\newblock
\APACrefYearMonthDay{1982}{4}{},
\newblock
\unskip
\newblock
\APACjournalVolNumPages{Sov. J. Nucl. Phys. (Engl. Transl.); (United
  States)}{35:4}{}{}.
\newblock
\begin{APACrefURL} \url{https://www.osti.gov/biblio/6558067} \end{APACrefURL}
\PrintBackRefs{\CurrentBib}

\bibitem [\protect \citeauthoryear {%
Witten%
}{%
Witten%
}{%
{\protect \APACyear {1984}}%
}]{%
Witten1984}
\APACinsertmetastar {%
Witten1984}%
\begin{APACrefauthors}%
Witten, E.%
\end{APACrefauthors}%
\unskip\
\newblock
\APACrefYearMonthDay{1984}{Jul}{},
\newblock
\unskip
\newblock
\APACjournalVolNumPages{Phys. Rev. D}{30}{}{272--285}.
\newblock
\begin{APACrefURL} \url{https://link.aps.org/doi/10.1103/PhysRevD.30.272}
  \end{APACrefURL}
\PrintBackRefs{\CurrentBib}

\bibitem [\protect \citeauthoryear {%
Yang%
, Ping%
\BCBL {}\ \BBA {} Segovia%
}{%
Yang%
\ \protect \BOthers {.}}{%
{\protect \APACyear {2020}}%
}]{%
Yang2020}
\APACinsertmetastar {%
Yang2020}%
\begin{APACrefauthors}%
Yang, G.%
, Ping, J.%
\BCBL {}\ \BBA {} Segovia, J.%
\end{APACrefauthors}%
\unskip\
\newblock
\APACrefYearMonthDay{2020}{}{},
\newblock
\unskip
\newblock
\APACjournalVolNumPages{Symmetry}{12}{11}{}.
\newblock
\begin{APACrefURL} \url{https://www.mdpi.com/2073-8994/12/11/1869}
  \end{APACrefURL}
\PrintBackRefs{\CurrentBib}

\end{thebibliography}

\end{document}